\documentclass[twocolumn,preprintnumbers,amsmath,amssymb,nofootinbib,superscriptaddress]{revtex4-1}
\usepackage{hyperref}
\usepackage{graphicx}% Include figure files
\usepackage{color}
\usepackage{xcolor}
\newcommand{\nonumbo}{\nonumber \\ && }
\newcommand{\backo}{\!\!\!\!\!\!\!\!\!\!}

\newcommand{\be}{\begin{equation}}  
\newcommand{\ee}{\end{equation}}   
\newcommand{\ba}{\begin{array}}  
\newcommand{\ea}{\end{array}}

\usepackage{diagbox}

\definecolor{rossoCP3}{cmyk}{0,.88,.77,.40}
\definecolor{blueRef}{rgb}{0.2,0.2,0.6}
\definecolor{blue}{rgb}{0,0.396,0.741}
\hypersetup{
	colorlinks, 
	bookmarksopen, 
	bookmarksnumbered,
	citecolor=blueRef, 		%color of links to bibliography
	linkcolor=rossoCP3,	%color of internal links
	urlcolor=rossoCP3,			%color of external links 
}

\newskip\humongous \humongous=0pt plus 1000pt minus 1000pt

\newif\ifdtup

\def\oldreffmt#1{\rlap{[#1]} \hbox to 2\parindent{}}

\def\figfmt#1{\rlap{Figure {#1}} \hbox to 1in{}}  
  
\def\beq{\begin{equation}}  
\def\eeq{\end{equation}}  
\def\bea{\begin{eqnarray}}  
\def\eea{\end{eqnarray}}  
\def\half{\frac{1}{2}}  
  
\def\bq{\begin{quote}}  
\def\eq{\end{quote}}

\def\half{\frac{1}{2}}       

\relax

\newdimen\tdim  
\tdim=\unitlength  
\def\bar{\overline}

\usepackage{color}

\usepackage{graphicx,graphics}

\usepackage{ulem}

\begin{document}
\preprint{FERMILAB-PUB-22-783-T}
%\preprint{}

\title {Renormalization Group for Non-minimal $\phi^2R$ Couplings
  \\  and Gravitational Contact Interactions}

%\vspace{0.5in}

\author{Dumitru Ghilencea}
\email{dumitru.ghilencea@cern.ch}
\affiliation{Department of Theoretical Physics, National Institute of Physics \\
and Nuclear Engineering (IFIN), Bucharest 077125, Romania}
\author{Christopher T. Hill}
\email{hill@fnal.gov}
\affiliation{Fermi National Accelerator Laboratory\\
P.O. Box 500, Batavia, Illinois 60510, USA\\$ $}

\date{\today}

\begin{abstract}
Theories of scalars and gravity, with an Einstein-Hilbert term and  non-minimal interactions, 
$M^2R/2 -\alpha\phi^2R/12 $, have graviton exchange induced contact interactions.
These modify the renormalization group, leading to a discrepancy between 
the conventional calculations in the Jordan frame that ignore this effect
(and are found to be incorrect),
and the Einstein frame in which $\alpha$ does not exist.
Thus, the calculation of quantum effects in the Jordan and Einstein frames does not generally commute
  with the transition from the Jordan to the Einstein frame.
In the Einstein frame, though $\alpha$ is absent,  for small steps in scale $\delta\mu/\mu$
infinitesimal contact terms $\sim \delta\alpha$ are induced, that are then
absorbed back into other couplings by the contact terms. 
This modifies the $\beta$-functions in the Einstein frame.  We show how
correct results can be obtained in a simple model by including this effect.
\end{abstract}

%\pacs{14.80.Bn,14.80.-j,14.80.-j,14.80.Da}
\maketitle

\section{Introduction}

Over the years there has been considerable interest in Brans-Dicke,
scalar-tensor, and scale or Weyl invariant theories.   
These  have in common  fundamental scalar fields, $\phi_i$, that couple to
gravity through non-minimal interactions, {$F(\phi_i)R$.   Unless forbidden by a symmetry (scale/Weyl),
  the Einstein term $M^2 R$ and the Planck mass, $M$, can co-exist with these non-minimal couplings,
  otherwise $M$ is generated dynamically by the VEV's of some of these  scalars \cite{authors}. }
When the theory is prescribed with non--minimal interactions we say it
is given in a ``Jordan frame.''

A key tool in the analysis of these models is the Weyl transformation, \cite{Weyl}.  This involves
a redefinition of the metric, $ g' =\Omega(\phi_i)g$, in which $g$ comingles 
with the scalars. $\Omega$ can be chosen to 
lead to a new effective theory, typically one that has a  pure Einstein-Hilbert action,
  $\sim M^2 R $, in which the non-minimal interactions 
  have been removed. This is called the ``Einstein frame.'' 
  Alternatively, one might use a Weyl transformation to partially 
  remove a subset of scalars from the non-minimal interactions $\sim M^2 R + F'(\phi_i)R$, 
  where $F'$ is optimized for some particular 
application.

It is {\it a priori}
unclear, however, how or whether the original Jordan frame  theory can be physically equivalent to the
Einstein frame form and how the Weyl transformation is compatible with a full quantum theory 
\cite{Duff}\cite{Ruf}.
Nonetheless, many authors consider this to be a valid transformation
and a symmetry of Weyl invariant theories, and many loop calculations permeate the literature
which attempt to exploit apparent simplifications offered by the Jordan frame.

It has been shown that any theory with non-minimal couplings and a Planck mass, $M$, contains
{\it contact terms} \cite{HillRoss}. These  are generated by the graviton exchange
amplitudes in tree approximation and they are therefore ${\cal{O}}( \hbar^0) $ and 
therefore classical. 
The contact terms occur because emission vertices from the non-minimal interaction are 
proportional to $q^2$ of the graviton, while the Feynman propagator
is proportional to $1/q^2$.  The cancellation of $q^2\times 1/q^2$ 
therefore leads to point-like interactions
that {\it must be included into
the effective action of the theory} at any given order of perturbation theory. 
 The result is that the non-minimal
interactions  disappear from the theory and Planck suppressed higher dimension operators 
appear with modified couplings.

The structural form of the theory when the contact
term interactions are included corresponds formally to
a Weyl transformation of the metric that takes the theory to the Einstein frame.
In the pure Einstein-Hilbert action there are no classical contact terms,
but at loop level they will be be generated, and must be removed as part of the
renormalization group.
However, by virtue of the contact terms, nowhere is a metric redefinition performed
(hence the issue of a Jacobian in the measure of the gravitational
path integral in going to the Einstein frame becomes moot).

This means that,  provided we are interested in the  theory  on mass scales below $M$, 
 the Jordan frame is an illusion and doesn't
 really exist physically. In the Jordan frame the contact terms are hidden, but they are
 always present.
Ergo, even though the action superficially appears to have non-minimal couplings, 
 it doesn't, and remains always in an Einstein frame. 
 
 Efforts to compute quantities, such as effective potentials (or equivalently,
 $\beta$-functions), in the Jordan frame, while ignoring the contact terms,
 will yield incorrect results. Nonetheless, though the non-minimal interactions
 are not present in the classical Einstein frame, they are regenerated
 by loops and the potentials and RG equations are modified by this effect.

 In a simple theory in the Jordan frame, where the non-minimal 
 coupling is $-\alpha \phi^2 R/12$,
 this raises the question of how to understand 
 the fate of $\alpha$?  With
 a single scalar with quartic and other interactions, $\lambda_i$, the usual ``naive'' calculation
 of a $\beta$-function in the Jordan frame (``naive'' means ignoring contact terms) yields the form:
 \bea
 \label{one1}
 \frac{\partial \alpha(\mu)}{\partial \ln(\mu)}= \beta_\alpha(\lambda_i) 
 \equiv (1-\alpha)\gamma_\alpha(\lambda_i)
 \eea
 where the factor $(1-\alpha)$ reflects the fact that when $M^2=0$ and  $\alpha=1$ (conformal limit)
 the kinetic term
 of $\phi$ disappears  and $\phi $ becomes static parameter).
 
 However, the contact terms (or a Weyl transformation) remove $\alpha$  
 and leave an Einstein frame with only the Einstein-Hilbert term, $M^2R$ and the $\phi$ couplings $\lambda_i'$
 (in what follows {\it primed couplings refer to the Einstein frame and un-primed to Jordan frame}).
 This means that  $N$ couplings, $(\alpha, \lambda_i)$, in the Jordan 
 frame have become  $N-1$
 couplings, $(\lambda_i')$, in the Einstein frame.  Therefore any physical 
 meaning ascribed to $\alpha$ or
 the $\beta_\alpha$ function is apparently lost.

 Three Feynman diagrams (shown below as D1, D2, D3, in  Figs.(2,3,4))  contribute to $\beta_\alpha$
 in the Jordan frame.  One of them multiplies $\alpha$ in the Jordan frame (D3) and 
 yields the $(1-\alpha)$ factor in eq.(\ref{one1}). 
 But even with $\alpha=0$ in the Einstein frame, two diagrams (D1 and D2) exist and
 reintroduce a perturbative $\delta\alpha$, for a small step in scale $\delta\mu/\mu$.
 This is then removed by the contact terms, but leads to  correction terms in the 
 renormalization of the $\lambda_i'$. 
 We are therefore sensitive to the same scale breaking information
 in the Einstein frame that one has in the Jordan frame, 
 which is encoded into $\gamma_\alpha$. 
 This does not, however, imply that the resulting 
 calculations in the Jordan and Einstein frames are then consistent!
 We explicitly demonstrate the inconsistency through calculation of
 effective potentials (the RG equations  of the couplings can always be
 read off from the effective potentials).

 If we stayed in the Jordan frame, with nonzero $\alpha$, and naively
 computed the same effective potential (``naively'' means ignoring contact terms),
 into an Einstein frame, we would obtain a different result.
 The difference  is a term proportional to $\alpha$ in the Jordan frame.
 Equivalently,  going initially to the  Einstein frame 
 and running with the RG, {\it does not commute}
 with running initially in the Jordan frame and subsequently going to the Einstein frame!

 We turn presently to a brief discussion of contact terms in general
 and review a simple toy model from \cite{HillRoss} that 
 is structurally similar to the gravitational case.
 We then summarize gravitational contact terms (and refer the reader to \cite{HillRoss}\cite{staro}
 for details and applications). We then exemplify the Einstein frame
 renormalization group compared to the naive Jordan frame result, which ignores contact terms,
 and illustrate the discrepancy.

\section{Contact Interactions}

Generally speaking ``contact interactions'' are point-like operators that are
generated in the effective action of the theory in perturbation theory.
They may arise in the UV from ultra-heavy fields that are integrated out, such as the Fermi
weak interaction that arises from integrating out the heavy $W$-boson. 
They may also arise in the IR when a vertex in the theory is proportional to $q^2$ and
cancels against a $1/q^2$ propagator. The gravitational contact term we discuss presently is of the IR form.

\subsection{Contact Terms in Non-gravitational Physics}

Contact terms arise in a number of
phenomena.   Diagrammatically they can arise in the IR when a
vertex for the emission of, e.g., a massless  quantum, of momentum $q_\mu$,
is proportional to $q^2$.  This vertex then cancels the $1/q^2$
from a massless propagator when the quantum is exchanged. This $q^2/q^2$
cancellation leads to an
effective point-like operator from an otherwise single-particle reducible diagram.

For example, in electroweak physics a
vertex correction by a $W$-boson to a massless gluon emission 
induces a quark flavor changing operator, e.g.,
describing $s\rightarrow d$+gluon, where $s(d)$ is a strange (down) quark.
This has the form of a local operator \cite{SVZ}\cite{penguins}:
\bea
\label{one}
g\kappa \bar{s}\gamma_\mu T^A d_L D_\nu G^{A\mu\nu}
\eea
where $G^{A\mu\nu}$
is the color octet
gluon field strength and $\kappa \propto
G_{Fermi}$.

This implies a vertex 
for an emitted gluon of 4-momentum $q$ and polarization and color, $\epsilon^{A\mu}$,
of the form
$g\kappa\bar{s}\gamma_\mu T^A d_L  \epsilon^{A\mu}\times q^2 +...$.  However,
the gluon propagates, $\sim 1/q^2$, and couples to a quark current 
$\sim g\epsilon^{A\mu}\bar{q}\gamma_\mu T^A q$.
This results in a contact term:
\bea
\label{localop}&&
g^2 \kappa\left(\frac{q^2}{q^2}\right) \bar{s}\gamma^\mu T^A d_L \bar{q}\gamma_\mu T^A q
\;=\;
g^2 \kappa\bar{s}\gamma^\mu T^A d_L\bar{q}\gamma_\mu T^A q
\nonumbo
\eea
The result is a 4-body local operator 
which mediates electroweak transitions
between, e.g., kaons and pions \cite{SVZ}, also
known as ``penguin diagrams'' \cite{penguins}.
Note the we can rigorously obtain the contact term result
by use of the gluon field equation within the operator
of eq.(\ref{one}), 
\bea
D_\nu G^{A\mu\nu}= g\bar{q}\gamma^\mu T^A q.
\eea
This is justified as operators that vanish by equations of motion, 
known as ``null operators,'' will generally have
gauge non-invariant anomalous dimensions and are unphysical 
\cite{Deans}.

Another example occurs in the case of
a cosmic axion, described by an oscillating  classical 
field,  $\theta(t)=\theta_0\cos(m_at)$, 
interacting with a magnetic moment, $ \vec{\mu}(x)\cdot \vec{B}$, through the
electromagnetic anomaly $ \kappa \theta(t) \vec{E}\cdot \vec{B} $. A static magnetic moment 
emits a virtual space-like photon of momentum $(0,\vec{q})$. The anomaly 
absorbs the  virtual photon and emits an on-shell photon of polarization $\vec{\epsilon}$,
inheriting energy $\sim m_a$ from the cosmic axion.
The Feynman diagram, with the exchanged virtual
photon, yields an amplitude, 
$\propto (\theta_0\mu^i\epsilon_{ijk}q^j)(1/\vec{q}^{\;2}) (\kappa \epsilon^{k\ell h}q_\ell m_a\epsilon_h)
\sim  (\kappa\theta_0 m_a\vec{q}^{\;2}/\vec{q}^{\;2}) \vec{\mu}\cdot \vec{\epsilon}$. 
The $ \vec{q}^{\;2}$ factor then 
cancels the $1/\vec{q}^{\;2}$ in the photon propagator,
resulting in a contact term which is an induced, parity violating,
oscillating electric dipole interaction:
$\sim \kappa \theta (t) \vec{\mu}\cdot \vec{E}
$. This results in 
cosmic axion induced {\it electric dipole} radiation from
any magnet, including an electron \cite{CTHa}.

\subsection{Illustrative Toy Model of Contact Terms}

To illustrate the general IR contact term phenomenon,
consider a single massless real scalar field $\phi $
and operators $A$ and $B,$ which can be functions of other fields, with the 
action given by:
\bea
S=\int \frac{1}{2}\partial \phi \partial \phi -A\partial ^{2}\phi -B\phi 
\label{toy0}
\eea
where $A$ and $B$ are functions of other fields.

Here  $\phi $ has a propagator ${i}/{q^{2}}$, but the vertex of a diagram
involving
$A$ has a factor of $\partial^2 \sim -q^{2}$. This yields a point-like interaction, $\sim
q^{2}\times ({i}/q^{2})$,
in a single particle exchange of $\phi $, and therefore implies contact terms.

At lowest order in perturbation theory consider the diagram with $\phi$
exchange in Fig.(1).  This involves two time-ordered products
of interaction operators: 
\bea
&& 
 {T} \;\;  i\!\!\int \! A\partial ^{2}\phi \times i\!\!\int \! B\phi 
 \rightarrow \frac{iq^{2}}{q^{2}}\
 AB\;\;\;=\; i\int \! AB,
\nonumbo 
\frac{1}{2}{T} \;\;  i\!\!\int \! A\partial ^{2}\phi \times
i\!\!\int \! A\partial ^{2}\phi 
 \rightarrow 
\frac{i(q^{2})^{2}}{2q^{2}}A^2   
\!=\!\frac{i}{2}%\!\!
\int \! A\partial ^{2}A.
\nonumbo
\eea
where $d^4x$ is understood in the integrals
and the $\half$ factor in the  $A\partial^2A$ term comes from the second order
in the expansion of the path integral $\exp (i\int A\partial^2 \phi)$. Note that
we also produce a non-local  interaction $ -{i}B^2/{2q^{2}}.$ 

\begin{figure}[t!]
\vspace{0.0 in}
	\hspace*{0.1in}\includegraphics[width=0.75\textwidth]{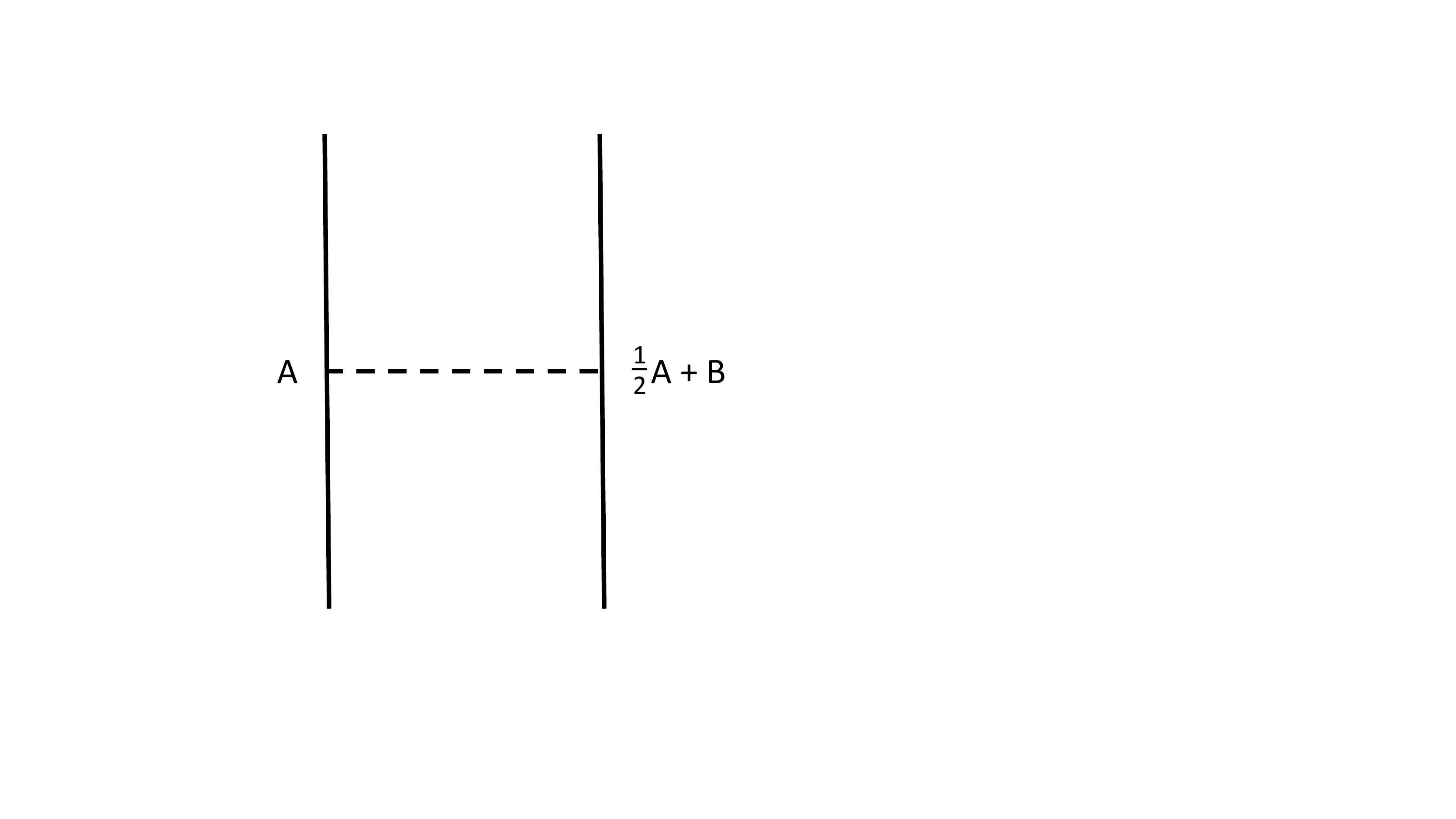}
	\vspace{-0.5 in}
	\caption{Contact terms in the toy model are
	generated by diagrams with exchange of $\phi$ (dashed). In gravity,
	with non-minimal term $\sim \int\! \sqrt{-g} F\left(\phi\right) R$ 
	and matter field Lagrangian $\sim \int\! \sqrt{-g} L\left(\phi\right)$
	then $A$ is replaced by $F(\phi)$ and $B$ is replaced by $L(\phi)$, and the dashed line
	is a graviton propagator. }
\end{figure}
We thus see that we have
diagrammatically obtained a local effective action:
\bea
S=\int \frac{1}{2}\partial \phi \partial \phi +\frac{1}{2}A\partial
^{2}A+AB +\makebox{ {long distance}}.
\eea
Of course, we can see this straightforwardly by
``solving the theory,'' by defining a shifted field:
\bea
\phi =\phi^{\prime } -\frac{1}{\partial ^{2}}\left( \partial ^{2}A+B\right).
\eea
Substituting this into the action $S$  and integrating by parts yields:
\bea
\label{eff1}
S =\int \frac{1}{2}\partial \phi ^{\prime }\partial \phi ^{\prime }+\frac{%
1}{2}A\partial ^{2}A+AB+\frac{1}{2}B\frac{1}{\partial ^{2}}B.
\eea
An equivalent effective local action that describes both
short and large distance is then,
\bea
\label{efftoy}
S=\int \frac{1}{2}\partial \phi \partial \phi +\frac{1}{2}%
A\partial ^{2}A+AB-{B \phi }.
\eea
The contact terms have become point-like components of the effective action,
while the remaining
{long} distance effects are produced by the usual massless $\phi $ exchange.  Note that the
derivatively coupled
operator $A$ has no long distance interactions {due to} $\phi $ exchange. 
Moreover, in the
effective action of
eq.(\ref{efftoy}) we have implicitly ``integrated out'' the $A\partial ^{2}\phi $, which is no
longer part of the action and is replaced
by new operators $\frac{1}{2}A\partial ^{2}A+AB$.
We will see that this is exactly what happens with gravity, where
the $A\partial^{2}\phi$ term is schematically the non-minimal $F(\phi)R(g)\sim F(\phi) \partial^2 h$
term in a weak field expansion of gravity $g =\eta+h$.

One can also adapt the use of equations of motion to obtain eq.(\ref{efftoy})
from the action eq.(\ref{toy0}) but
this requires care. For example, the insertion of the $\phi$ equation of motion,
into $A\partial^2\phi$ correctly gives the $AB$ term but misses the 
factor of $1/2$ in the $A\partial^2A$ term.  We can therefore do a trick of
defining a
``modified equation of motion,'' where we supply a factor of $1/2$ on the term  $A\partial^2A$,
e.g. substitute:
\bea
\label{subst}
\partial ^{2}\phi=-\partial ^{2}A-B \rightarrow  -\frac{1}{2}\partial ^{2}A-B
\eea
in place of the $ \partial ^{2}\phi$ in the second term of eq.((\ref{toy0}))
to obtain eq.(\ref{efftoy}).

\section{Gravitational Contact Terms}

Consider a general theory involving scalar fields $\phi _{i},$
an Einstein-Hilbert {term and} a non-minimal interaction:
\bea
\label{action00}
&&\backo\;\;
S=\int \sqrt{-g}\left(\frac{1}{2}M^{2}R\left( g_{\mu \nu }\right) +\frac{1}{2%
}F\left( \phi _{i}\right) R\left( g_{\mu \nu }\right) +L\left( \phi
_{i}\right) \right)
\nonumbo
= S_1 + S_2 +S_3
\eea
where we use the 
metric signature and curvature tensor conventions of \cite{CCJ}.
 $S_{1}$ is the kinetic term of gravitons
\bea
S_{1} &=& \frac{1}{2}M^{2}\int \sqrt{-g}R
\eea
and becomes the Fierz-Pauli action in a weak field expansion.

$S_{2}$ is the non-minimal interaction, and takes the form:
\bea
S_{2}=\frac{1}{2}\int \sqrt{-g}F\left( \phi _{i}\right) R\left( g_{\mu \nu }\right)
\eea
{where $F$ is polynomial in fields.}

$S_{3}$ is the matter action with couplings to the gravitational weak
field:
\bea
S_{3}=\int \sqrt{-g}\;L\left( \phi _{i}\right) 
\eea
The Lagrangian takes the form
\bea
L\left( \phi_{i}\right)  = \frac{1}{2}g^{\mu \nu }\partial _{\mu }\phi_i \partial _{\nu }\phi_i-W(\phi_i)
\eea
with potential $ W(\phi_i) $.
The matter Lagrangian has stress tensor and stress tensor trace:
\bea
T_{\mu \nu }&=&\partial _{\mu }\phi_i \partial _{\nu }\phi_i -g_{\mu \nu }\left( 
\frac{1}{2}g^{\rho \sigma }\partial _{\rho }\phi_i \partial _{\sigma }\phi_i
-W(\phi_i )\right) .
\nonumber \\
T&=&-\partial ^{\sigma }\phi_i \partial _{\sigma
}\phi_i +4W(\phi_i)
\eea
There are then three ways to obtain the contact term:

 \vspace{0.2in}
\noindent
 (1) GRAVITON EXCHANGE CONTACT TERM
 
 \vspace{0.2in}
 \noindent

 In reference \cite{HillRoss} the corresponding Feynman diagrams of Figure 1
 are evaluated, arising from single graviton exchange between the interaction terms.  

We treat the theory perturbatively, expanding around flat space. {Hence} we
linearize gravity with a weak field $h_{\mu \upsilon }$:
{\bea
\label{linear}
g_{\mu \upsilon }\approx \eta _{\mu \upsilon }+\frac{h_{\mu \upsilon }}{%
M}.
\eea}
The scalar curvature is then:
\bea
R &=& R_{1}+R_{2}
\nonumber \\
M R_{1}&=&\biggl( \partial^{2}h\ -\partial^{\mu}\partial ^{\nu}h_{\mu \nu }\biggr) 
\nonumber \\
M^{2}R_{2}&=&  -\frac{3}{4}\partial ^{\rho }h^{\mu
\nu }\partial_{\rho}h_{\mu \nu }
-\frac{1}{2}h^{\mu
\nu }\partial ^{2}h_{\mu \nu }+... 
\eea
(see \cite{HillRoss} for the complete expression for $R_2$).
 $S_{1}$ then becomes:
\bea &&
\label{FP0}
S_{1} = \frac{1}{2}M^{2}\int \sqrt{-g}R
= \frac{1}{2}%
M^{2}\int\!\! \left( R_1+R_{2}+\frac{1}{2}\frac{h}{M}R_{1}\right) 
\nonumbo
= \frac{1}{2}\int h^{\mu \nu }\biggl( \frac{1}{4}\partial ^{2}\eta _{\mu \nu
}\eta _{\rho \sigma }-\frac{1}{4}\partial ^{2}\eta _{\mu \rho }\eta _{\nu
\sigma }
\nonumbo \qquad\qquad
-\frac{1}{2}\partial _{\rho }\partial _{\sigma }\eta _{\mu \nu }+%
\frac{1}{2}\partial _{\mu }\partial _{\rho }\eta _{\nu \sigma }\biggr)
h^{\rho \sigma }.
\eea
 Note that the leading term, $\int R_{1}$,
is a total divergence and is therefore zero in the Einstein-Hilbert
action, and what remains of eq.(\ref{FP0})
is the Fierz-Pauli action.   This is key to the origin of the contact terms.
The non-minimal interaction, $S_{2}$, then takes the leading form:
\bea &&
S_{2}=\frac{1}{2}\int \!\!\sqrt{-g}\; F\left( \phi\right) R\left( g\right)
\rightarrow \frac{1}{2}\int \!\!F\left( \phi\right) R_1 \left( g\right)
\nonumbo
\qquad
=
\int\frac{1}{2M}F\left( \phi\right) 
\Pi
^{\mu \nu }h_{\mu \nu } 
\eea
where it is useful to introduce the transverse derivative,
\bea
\Pi ^{\mu \nu }= \partial ^{2}\eta ^{\mu \nu }\ -\partial
^{\mu }\partial ^{\nu } .
\eea 
$S_{2}$ involves derivatives, since $R_1$ is now active,  
and is the analogue of the $A\partial^2\phi$ term in eq.(\ref{toy0}).
It will therefore generate
contact terms in the gravitational potential due to single graviton
exchange. $S_3$ is the analogue of the $B\phi$ term in eq.(\ref{toy0}), and 
this situation will closely parallel the toy model.

In \cite{HillRoss} we developed the graviton propagator, following
the nice lecture notes of
Donoghue {\it et. al.,} \cite{Donoghue}.
We remark that we found a particularly
useful gauge choice,
\bea
\label{wgauge}
\partial_\mu h^{\mu \nu} = w \partial^\nu h
\eea
where $w$ defines a single parameter family
of gauges.
The familiar De Donder gauge corresponds to $w=\frac{1}{2}$, 
while the  choice $w=\frac{1}{4}$ is particularly natural in
this application, and the gauge invariance
of the result is verified by the $w$-independence
(we verify the Newtonian potential from graviton
exchange between static masses in $w$ gauge; see \cite{HillRoss}).

We can then compute single graviton exchange between the interaction
terms of the theory.
A diagram with a single
$S_2$ vertex and single $S_3$ vertex is the analogue of $AB$ in the toy model and yields:
\bea
-i\left\langle  T\;S_{2}S_{3}\right\rangle=
\int d^{4}x\; \frac{F\left( \phi _{i} \right) }{2M^{2}}%
T( \phi _{i}) 
\eea
Also we have 
the pair $\left\langle S_{2}S_{2}\right\rangle $
which corresponds to $\frac{1}{2} A\partial^2 A$ in the toy model and yields:
\bea 
-i{\left\langle T\; S_{2}S_{2}\right\rangle}&=&-\int d^{4}x\;  \frac{3}{4M^{2}}\; F\left(
\phi _{i} \right) \!\partial ^{2}F\left( \phi _{i}\right) 
\nonumbo
\eea
The action becomes
\bea
 &&
S = S_1+S_3+ S_{CT}
\eea
where
\bea
\label{CT}
&&
S_{CT}= 
\int d^{4}x \biggl( 
-\frac{3}{4M^{2}} F \partial ^{2}F 
+\frac{1 }{2M^{2}}FT
\biggr)
\eea
Note the sign of the $F\partial ^{2}F$ is opposite (repulsive) to that of
the toy model $A\partial^2 A$.

\vspace{0.2in}
\noindent
(2) WEYL TRANSFORMATION 
\vspace{0.2in}
 \noindent

Define
\bea
\Omega ^{2}=\left( 1+\frac{F\left( \phi_i \right) }{M^{2}}\right)
\eea
and 
perform a Weyl transformation on the metric:
\bea
\label{metricweyl}
&&
g_{\mu \nu }(x)\rightarrow \Omega ^{-2}g_{\mu \nu }(x)
\qquad
g^{\mu\nu }(x)\rightarrow \Omega ^{2}g^{\mu \nu }(x)
\nonumbo
\sqrt{-g}\rightarrow\sqrt{-g}\Omega^{-4}
\nonumbo
R(g)\rightarrow\Omega ^{2}R(g)+6\Omega ^{3}
% D\partial
\,\Box\,\Omega ^{-1}
\eea
and the action of {eq.(\ref{action00})} becomes:
\bea &&
S\rightarrow\int \sqrt{-g}\biggl(\frac{1}{2}M^{2}
R\left( g\right) 
\nonumbo
-3M^{2}\partial_\mu \left( 1+\frac{F}{
M^{2}}\right)^{1/2}\!\!\partial^\mu 
\left( 1+\frac{F}{M^{2}}\right)
^{-1/2} 
\nonumbo
 +\frac{1}{2}\left(1+\frac{F}{M^{2}}\right)^{-1}\!\!\!
\partial _{\mu }\phi_i \partial^{\nu }\phi^i 
-\left( 1+\frac{F}{M^{2}}\right)^{-2}\!\!\!W(\phi_i)\biggr)
\nonumbo
\eea
Keeping terms to $O({1/M^2})$ and integrating by parts we have:
\bea && \backo
S=S_{1}+S_3
\nonumbo \backo
+\int d^4x\bigg(-\frac{3
F\left( \phi _{i}\right)\partial ^{2}F\left( \phi _{i}\right)}{4M^{2}}
+\frac{F\left( \phi _{i}\right)T\left( \phi _{i}\right)}{2M^{2}}\bigg) 
\eea
The Weyl transformed action is identically consistent with the contact terms
of eq.(\ref{CT}) above,
to first order in $1/M^2$.

Hence, contact terms arise in gravity with non-minimal
couplings to scalar fields due to graviton
exchange. {Their form is equivalent to
a Weyl redefinition of the theory to the  Einstein frame
action and reinforces their role as induced
components of the effective action.}  Hence any theory
with a non-minimal interaction $\sim F(\phi)R$ will lead
to  contact terms at order $1/M^2$. 

The Weyl transformation is non-perturbative.
It is technically simpler  than the gravitational
potential
calculation, and it confirms the tricky normalization factors and phases in
the graviton
exchange calculation. As the Weyl transformation makes no reference to a gauge
choice, a calculation of the the contact terms in other gauges
should yield the
equivalent results. To check the invariance we turn now to a calculation
in an alternative gauge which sheds further light on the origin of their structure.

 \vspace{0.2in}
\noindent
(3)  USE OF MODIFIED $R$ EQUATION OF MOTION
\vspace{0.2in}
 \noindent

 A  trick can be used to simplify the calculations below.
The Einstein equation with the non-minimal term is:
\bea &&
M^{2} 
G_{\alpha \beta }=-T_{\alpha \beta }-\nabla_\mu(\nabla_\nu F(\phi_i)) + g_{\mu\nu} \nabla^2 F(\phi_i)
\nonumbo
M^2 R=  T - 3 \nabla^2 F(\phi_i)
\eea
To use a modified ``equation of motion'' we first supply a factor of $1/2$
in the last term which is the analogue of the $\partial^2 A$ term
as in eq.(\ref{subst}):
\bea 
\label{RE0} &&
 R'=\frac{1}{M^2} \biggl( T - \frac{3}{2} \nabla^2 F(\phi_i) \biggr)
\eea
Then
substitute $R'$ for $R$  in the non-minimal term $F R$ of the 
{action of eq.(\ref{action00})}
\bea &&\backo \;\;
S\rightarrow\int \sqrt{-g}\left(\frac{1}{2}M^{2}R\left( g_{\mu \nu }\right) +\frac{1}{2%
}F\left( \phi _{i}\right) R'\left( g_{\mu \nu }\right) +L\left( \phi
_{i}\right) \right)
\nonumbo
 \qquad =S_{1}+S_3
\nonumbo
\qquad
+
\int d^4x\bigg(-\frac{3
F\left( \phi _{i}\right)\partial ^{2}F\left( \phi _{i}\right)}{4M^{2}}
+\frac{F\left( \phi _{i}\right)T\left( \phi _{i}\right)}{2M^{2}}\bigg)
\nonumbo
\eea
In the RG calculation we will only need the exact equation of motion 
for $R$ in the Einstein frame (without the pseudo $-3\nabla^2F/2$ term), 
so this ambiguity does not arise.

\section{Contact Term in a Simple Model}

Consider the following action for a single real scalar field $\phi$:
\bea
\label{simp}
&&\backo
S_{Jordan}=\int \sqrt{-g}\biggl( \frac{1}{2}\partial\phi\partial\phi
-\frac{\lambda _{1}}{4}\phi ^{4}-
\frac{\lambda _{2}}{12M^{2}}\phi ^{6}
\nonumbo
-\frac{\lambda_3 }{12M^{2}}\phi ^{2}\partial\phi\partial\phi
-\frac{\alpha}{12}\phi ^{2}R
+\frac{1}{2}M^{2}R
\biggr) 
\eea
$\partial^2\phi\partial^2\phi$ terms can be dealt with by  adding a $k^4$ term in
the kinetic term of $\hat\phi$ .  It is then possible to see this does not modify the present operator
basis in the loops, and it would be equivalent to using the equations of motion
$\partial^2\phi = \lambda_1\phi^3$ which can be absorbed into a redefinition of $\lambda_2$. 

This is the most general action for $\phi$ with a $Z_2$ symmetry $\phi\rightarrow -\phi$
valid to O($M^{-2}$) with Einstein gravity and assuming $\phi$ is massless, $m^2=0$.
We will study this model to leading order $1/M^2$ and one loop, ${\cal{O}}\hbar$.
Hence we do not include a term $\phi^4R/M^{2}$ since, after use of equations of motion, $R\sim M^{-2}$
such a term would enter the physics at O($M^{-4}$).  Also note that, by integration by parts, 
$\int \phi^2 \partial^2\phi^2=(-4) \int \phi^2 (\partial \phi)^2$
and
$\int \phi^3 \partial^2\phi=-3\int \phi^2 (\partial\phi)^2$.

The matter Lagrangian has stress tensor and stress tensor trace
\bea &&
T_{\mu \nu }=\biggl( 1- \frac{\lambda_3\phi^2 }{6M^{2}}\biggr)\partial _{\mu }\phi \partial _{\nu }\phi -g_{\mu \nu }\left( 
\frac{1}{2}(\partial\phi)^2-W(\phi)\right) .
\nonumbo
T=-(\partial\phi)^2 +4W(\phi)
\nonumbo
W=\frac{\lambda _{1}}{4}\phi ^{4}+
\frac{\lambda _{2}}{12M^{2}}\phi ^{6}+\frac{\lambda_3 }{12M^{2}}\phi ^{2}(\partial\phi)^2
\eea
and we have
\bea
F =-\frac{\alpha}{6}\phi^2
\eea
which leads to the contact terms:
\bea&&
-\frac{3F\partial ^{2}F}{4M^{2}}
=\frac{\alpha^2(\phi\partial\phi)^2}{12M^{2}}
\nonumbo
\frac{FT}{2M^{2}}
=\frac{\alpha\phi^2 }{12M^{2}}\biggl((\partial\phi)^2 -\lambda _{1}\phi ^{4}
\biggr)
\eea

Therefore, the
effect of single graviton exchange to eq.(\ref{simp})
yields the Einstein frame action to order $M^{-2}$:
\bea &&
\label{simp2}
S_{Einstein}=\int \sqrt{-g}\biggl( \frac{1}{2}\partial\phi\partial\phi
-\frac{\lambda' _{1}}{4}\phi ^{4}-
\frac{\lambda' _{2}}{12M^{2}}\phi ^{6}
\nonumbo
-\frac{\lambda_3' }{12M^{2}}\phi ^{2}\partial\phi\partial\phi
+\frac{1}{2}M^{2}R
\biggr) 
\eea
where:
\bea \label{stuff0}
&&
\lambda_3' =\lambda_3 -\alpha-\alpha^{2}
\nonumbo
\lambda_1' = \lambda_1 
\nonumbo
\lambda_2'=\lambda_2+\alpha\lambda_1
\eea
Thus, the Planck suppressed  terms in Jordan frame (of couplings
  $\lambda_{2,3}$) that are usually ignored to  a leading approximation,
  are actually of same order ($1/M^2$) to the non-minimal term ($\phi^2 R$)
 when written in the Einstein frame.
We see that to first order in $M^{-2}$ in $S_{Einstein}$
we have three interaction terms,  though the original
action $S_{Jordan}$ displayed four interaction terms. In the latter
action
we see that $\alpha$ has disappeared having been absorbed into
redefining the primed coupling constants.
This indicates that the non-minimal term in $S_{Jordan}$ with coupling 
$\alpha$ is unphysical.

\subsection{Effective Action}

To compute the effective action for a classical background field
$\phi_0$ we
expand the action with a shifted field,
\bea
\phi\rightarrow \phi_0+\sqrt{\hbar}\hat{\phi}
\eea
to O$(\hat\phi^2)$. We will integrate out the quantum fluctuations
and can therefore
drop terms odd in $\hat{\phi}$, so we have:\footnote{This corresponds to using a source 
term, $J\phi$, to compute $S(J)$, then performing a  Legendre transformation 
to obtain the effective action as a function
of $\phi_0=\delta S(J)/\delta J$ as in \cite{CW}.}
\bea 
\label{}
&&\backo
S_{Einstein}
=
\nonumbo\backo
\int \sqrt{-g} \biggl(\frac{1}{2} \partial {\phi}_0 \partial{\phi}_0 +\frac{1}{2} \partial \hat{\phi} \partial\hat{\phi}
-\half B_0 \hat\phi^2 
-V_0
+\frac{1}{2}M^{2}R
 \nonumbo
 -\frac{\lambda_3' }{12M^2} (\hat{\phi}+\phi_0) ^{2}(\partial(\hat\phi+\phi_0))^2
\biggr) 
\eea
where
\bea &&
\label{unprimed}
B_0=\biggl({3\lambda'_1}\phi_0^{2}
+\frac{5\lambda' _{2}}{2M^2}\phi_0^4   \biggr)
\nonumbo
V_0=\frac{\lambda' _{1}}{4}\phi_0^{4}
+\frac{\lambda' _{2}}{12M^{2}}{\phi_0 ^{6}}
\eea

To treat  the $\lambda_3' $ term in the shifted fields 
we use integration by parts and the identity
$2\phi\partial\phi =\partial(\phi^2)$. We  obtain the terms
to O$\hat{\phi}^2$:
\bea &&
\label{redux}
\int (\hat{\phi}+\phi_0) ^{2}(\partial(\hat\phi+\phi_0))^2
\nonumbo
\rightarrow
\int 
\phi_0 ^{2} (\partial \phi_0)^2
- \phi_0^{2} \hat\phi\partial^2\hat\phi
+\hat\phi ^{2} \biggl(
\partial\phi_0 \partial\phi_0
-\half\partial^2(\phi^2_0) \biggr)
\nonumbo
\eea
For the O$(1/M^2)$ terms we can then use the equation of motion: 
\bea
\phi_0 ^{2} \hat\phi\partial^2\hat\phi\approx
- B_0\phi_0 ^{2}\hat{\phi}^2\approx -3\lambda'_1\phi_0^{4}\hat{\phi}^2
\eea
This is a bit tricky, but can be seen by consideration of Feynman
diagrams that contribute $\ln(\Lambda^2)$.\footnote{If we restrict ourselves to
constant $\phi_0$, as in Coleman-Weinberg \cite{CW}
we can considerably simplify the analysis and
just drop any terms with $\partial\phi_0$ and eq.(\ref{redux}) 
becomes $\phi_0^2(\partial\hat{\phi})^2$ and can be absorbed into
a wave-function renormalization of $\hat{\phi}$; this yields the result we obtain below when 
$\partial\phi_0=0$.}
Then we have:
\bea 
\label{E0}
&&\backo
S_{Einstein}
= S_c+\hbar S_q
\nonumbo\backo
S_c=\int \sqrt{-g} \biggl(\frac{1}{2} \partial {\phi}_0 \partial{\phi}_0  
-V'+\frac{1}{2}M^{2}R
\biggr) 
\nonumbo\backo
S_q=\int \sqrt{-g} \biggl( \frac{1}{2} \partial \hat{\phi} \partial\hat{\phi}
-\half B' \hat\phi^2 
\biggr) 
\eea
where,
\bea &&
B'
={3\lambda'_1}\phi_0^{2}
+\frac{5\lambda' _{2}}{2M^2}\phi_0^4 
\nonumbo
\qquad
+\frac{\lambda_3'}{6M^2}
\biggl( 3\lambda'_1\phi_0 ^{4}+(\partial\phi_0)^2
-\half\partial^2(\phi^2_0) \biggr)
\nonumbo
V'=\frac{\lambda' _{1}}{4}\phi_0^{4}
+\frac{\lambda' _{2}}{12M^{2}}\phi_0 ^{6}
+\frac{\lambda_3' }{12M^2}\phi_0 ^{2} (\partial \phi_0)^2
\eea
First consider the non-curvature terms, with the flat Minkowski space
metric $g_{\mu\nu}=\eta_{\mu\nu}$.
We obtain the effective potential from the log of the path integral
 discussed in Appendix,  eq.(\ref{19}),
 \bea
\label{19c}
\Gamma_0
&=&
-\frac{1}{2}  B'^{2} L   +O \biggl(  \frac{B^{3}}{\Lambda^{2}} \biggr)    
\eea
and we define the log as (see Appendix):
\bea
L=\frac{1}{32\pi^{2}}\ln\frac{\Lambda^{2}}{\mu^2}
\eea
with a generic infrared cut-off mass scale $\mu$.
Hence, squaring $B'$, the resulting potential is
to ${\cal{O}}(1/M^2)$:
\bea &&
\label{20}\backo \!\!
\Gamma_0 =
-\biggl(
\frac{9\lambda'^2_1}{2} \phi_0^{4}
+\frac{15\lambda'_2\lambda'_1}{2M^{2}}\phi_0^6
\nonumbo
+\frac{\lambda_3'\lambda'_1}{2M^{2}}{\phi_0^2}
\biggl( 3\lambda'_1\phi_0 ^{4}+(\partial\phi_0)^2
-\half\partial^2(\phi^2_0) \biggr)
\biggr)   L
\nonumbo 
=
-\biggl(
\frac{9\lambda'^2_1}{2} \phi_0^{4}
+\frac{15\lambda'_2\lambda'_1+3\lambda_3'\lambda'^2_1}{2M^{2}}{\phi_0^6}
\nonumbo
 +
 \frac{3\lambda_3'\lambda'_1}{2M^2}{\phi_0^2}(\partial\phi_0)^2
\biggr)  L
\eea
where we integrated by parts the ${\phi_0^2}(\partial^2\phi^2_0) $ term.

\subsection{ Inclusion of a Gravitationally Induced \\
Contact Term in the Einstein Frame}

Consider a weak field approximation to gravity where the metric becomes:
\bea &&
\label{weak}
g_{\mu \upsilon }\approx \eta_{\mu \upsilon }+\frac{h_{\mu \upsilon}}{M},
\qquad g^{\mu \upsilon }\approx \eta^{\mu \upsilon }-\frac{h^{\mu \upsilon}}{M}
\nonumbo
\sqrt{-g}\approx 1+\frac{1}{2}\frac{h}{M}\;\;
\makebox{where}\;\;  h=\eta ^{\mu \upsilon }h_{\mu \upsilon }
\nonumbo
R=g^{\mu \beta }R_{\mu \beta }=\frac{1}{M}\left( \partial ^{2}h\ -\partial ^{\nu
}\partial _{\rho }h_{\nu }^{\rho }\right) +O(h^2)
\eea
We choose $w$-gauge (this gauge is developed in \cite{HillRoss}; $w=1/2$ corresponds to 
the familiar ``de Donder gauge,'' see
\cite{Donoghue}), 
\bea
\label{grav1}
\partial _{\alpha }h^{\alpha \beta }=w\partial
^{\beta }h, \qquad  R=(1-w)\partial ^{2}h/M
\eea
and up to linear terms in $h_{\mu\nu}\hat{\phi}^2/M$,  eq.(\ref{E0}) becomes
\bea &&
S_q\rightarrow \int \biggl( \half \partial_{\mu }\hat\phi\partial^{\mu }\hat\phi
+\frac{1}{4}\frac{h}{M}\partial_{\mu }\hat\phi\partial^{\mu }\hat\phi
-\frac{h^{\mu \upsilon }}{2M}  \partial_{\mu }\hat\phi\partial _{\nu }\hat\phi
\nonumbo \qquad 
-\half \biggl(1+\frac{1}{2}\frac{h}{M}\biggr) B'\hat\phi^{2} \biggr)
\label{aaa}
\eea

If we now include effects of gravity we see that
the action in eq.(\ref{aaa})
generates two diagrams of Figs.(2,3) 
that are linear in the curvature $R$ (other diagrams contribute
$\sqrt{-g}$ factors for the resulting potential). 
We use the Wick-rotated, Euclidean loop momentum with cut-off $\Lambda$
and we thus have for D1:
\bea &&
T \;\left( h^{\mu \upsilon
}\partial _{\mu }\hat\phi\partial _{\nu }\hat\phi \right)(\hat\phi^{2}) 
\nonumbo
=\left( 2h^{\mu \upsilon }\right) \int \frac{%
d^{4}\ell }{(2\pi )^{4}}\frac{i}{\left( \ell +q\right) ^{2}}\frac{i}{\left(
\ell \right) ^{2}}(\left( q+\ell \right) _{\mu }\left( \ell
\right) _{\nu })
\nonumbo
=i\frac{\left( 1+2w\right) }{3}\left( \partial ^{2}h\right) L
\eea
where the quadratic divergence is projected away by a factor $P_2$
as discussed in the Appendix.
For D2 we have:
\bea &&
T \;\left( h\eta^{\mu \upsilon
}\partial _{\mu }\hat\phi\partial _{\nu }\hat\phi \right)(\hat\phi^{2}) 
\nonumbo
= 6q^{2}h \int_0^1dx\int \frac{d^{4}\ell }{(2\pi )^{4}}\frac{1 
}{\ell ^{4}}x(1-x)
\nonumbo
=i\left( 2\partial ^{2}h\right) L
\eea
hence, the contributions to the potential are
\bea &&
\backo\backo 
\Gamma_{D1}
=i \frac{B'}{4M} \biggl\langle\;T \;
\biggl( h^{\mu \upsilon }\partial _{\mu }\hat\phi \partial_{\nu }\hat\phi \biggr)
 \biggl( \hat\phi^{2} \biggr)
 \biggr\rangle
 \nonumbo
 =-\frac{ \bigl( 1+2w \bigr) }{12M}B^\prime \partial ^{2}h L
\nonumbo
\backo\backo 
\Gamma_{D2}
= -i\frac{B'}{8M} \biggl\langle\;T \;
\biggl(  h \eta^{\mu\nu}\partial_{\mu }\hat\phi \partial_{\nu }\hat\phi\biggr)
 \biggl( \hat\phi^{2} \biggr)
 \biggr\rangle
 \nonumbo
 =\frac{B'}{4M}\partial^{2}h  L
\eea
The diagrams  
{$\langle T\; h B^\prime\hat{\phi}^2\;\; B^\prime\hat{\phi}^2 \rangle$}
 generate the covariant $\sqrt{-g}\; \Gamma $ terms and do
not lead to curvature.
Note that
\bea &&
\frac{B'}{4M} \partial^{2}h
-\frac{ ( 1+2w)B' }{12M} \partial ^{2}h 
=\frac{( 1-w )B'}{6M}\partial ^{2}h
{=\frac{B^\prime}{6} R}
\nonumbo
\eea
Hence we have the potential from $D1+D2$,
\bea
\label{D1D2}
&& \Gamma_\alpha\equiv
\Gamma_{D1}+\Gamma_{D2}
= \frac{1}{6}B'R \;L\approx\frac{\lambda'_1\phi_0^2}{2}R L
\eea

{We thus see there is a non-minimal  term $(\delta\alpha/12)\,\phi_0^2R$ in the
potential generated by the loops,  of
the form $\delta\alpha =6\lambda_1^\prime \delta L$,
 from the $3\lambda_1^\prime$ term in $B'$,
 where we only keep leading terms in $1/M^2$  since $R\sim 1/M^2$.}
 
 We remove this term by using the contact term.
To implement the contact term we
use, in eq.(\ref{D1D2}), the leading order $R$ equation of motion in the Einstein frame
from $S_c$,
\bea
\label{R00}&&
R=\frac{1}{M^2}  T=\frac{1}{M^2}\biggl(-(\partial\phi_0)^2 +\lambda' _{1}\phi_0 ^{4}\biggr)
\eea
Here we omit the  $\lambda_2$ term in $B'$ which is suppressed by $M^{-2}$,
and the ${\cal O}\hbar$,  $\lambda^2_{1}L$ term which would lead to a $L^2\sim \hbar^2$ contribution.
We then have
  \bea
 && \Gamma_\alpha
=\frac{1}{2}\lambda'_1\phi^2_0  R \;L
 \rightarrow -\frac{\lambda'_1}{2M^2} \phi_0^2(\partial\phi_0)^2L+
 \frac{\lambda'^2_1}{2M^2}\phi_0^6L 
 \nonumbo
 \eea
Therefore, combining all effects the effective action becomes our
final result:
\bea&&
\label{all}
\backo
S=\int \half(\partial\phi_0)^2 +\half M^2R - \Gamma(\phi_0),
\nonumbo
\backo
\Gamma_E(\phi_0)\equiv 
\frac{\lambda'_{1}}{4}\phi_0^{4}
+\frac{\lambda'_{2}}{12}%
\frac{\phi_0 ^{6}}{M^{2}}
+\frac{\lambda_3' }{12M^2}\phi_0 ^{2} (\partial \phi_0)^2
\nonumbo\backo
- \biggl(
\frac{9\lambda'^2_1}{2} \phi_0^{4}
+\frac{15{\lambda'_1\lambda'_2 }+3\lambda'^2_1\lambda_3' -[\lambda^{\prime\, 2}_1]}{2M^{2}}\phi_0^6
\nonumbo\backo
+
\frac{3\lambda_3'\lambda'_1}{2M^2}{\phi_0^2}(\partial\phi_0)^2
+\frac{\lambda'_1}{2M^2} \phi_0^2(\partial\phi_0)^2
\biggr) L
\eea
where the term in $[..]$ comes from the gravitational effects of D1 and D2.

\begin{figure}[t!]
\vspace{0 in}
	\hspace*{-0.1in}\includegraphics[width=0.5\textwidth]{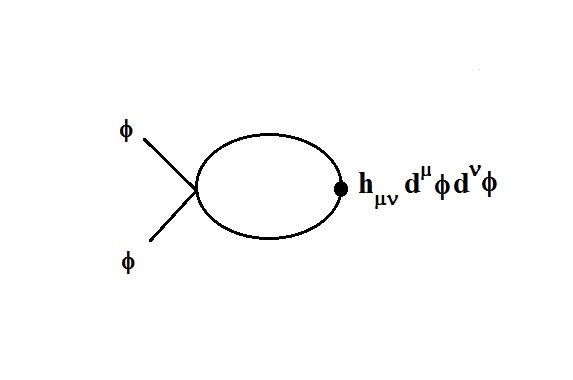}
	\vspace{-0.5 in}
	\caption{ Diagram D1. }
\end{figure}
\begin{figure}[t!]
\vspace{0 in}
	\hspace*{-0.1in}\includegraphics[width=0.5\textwidth]{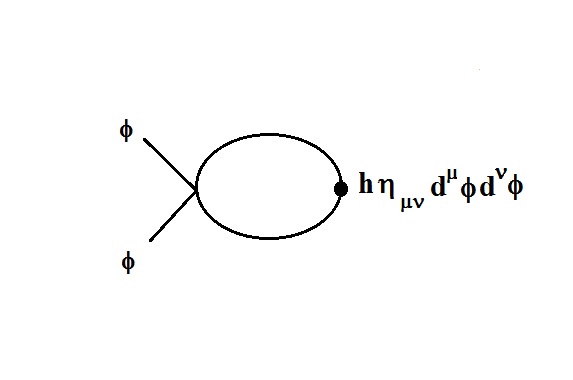}
	\vspace{-0.5 in}
	\caption{ Diagram D2.}
\end{figure}

From eq.(\ref{all})  we can read off the RG equations:
 \bea &&
 \label{RGF}
 D\lambda'_1=18\lambda_1^{\prime\, 2}
 \nonumbo
 D\lambda'_2=90\lambda'_1\lambda'_2+18\lambda_3'\lambda^{\prime\, 2}_1-6[\lambda^{\prime\, 2}_1]
  \nonumbo
 D\lambda_3'=18\lambda_3'\lambda_1'+6[\lambda'_1]
 \eea
where terms in $[...]$ comes from the ${\cal{O}}(\hbar)$ gravitationally induced contact term
 diagrams, D1 and D2.

The RG equations represent the differential inclusion of a loop
induced non-minimal coupling of eq.(\ref{D1D2}), $\delta\alpha=(6\lambda_1)\delta L$
occuring in the Einstein frame,
back into the $\beta_{\lambda_i}$ functions of the other couplings.  Thus it maintains
the reduction from $N$ to $N-1$ couplings (without $\alpha$) in the Einstein frame.
The only relevant parameters are the $\lambda_i$ in the Einstein frame
and they have a closed set of RG equations.
It is not surprising that such effects occur exclusively
in the Planck suppressed operators.  One might think
that these are not large effects, but they could be 
relevant when $\lambda_2 \gg \lambda_1$.  
The main point here is that the gravitational effects are 
present and must be included in Planck suppressed terms, 
but the calculation should be done in the Einstein frame
with implementation of the contact terms.

\subsection{Comparison to a Conventional Calculation of  $\beta_\alpha$ \\
in Jordan Frame Neglecting Contact Terms}

We now compute the effective potential in the Jordan frame where
we {\it naively neglect the contact term}, which is often seen in the literature.  
After evolving the theory in the Jordan frame we can then perform the Weyl transformation to compare
with the previous Einstein frame result.  As expected, these are found to be inconsistent.

In the Jordan action we  shift $\phi\rightarrow \phi_0+\sqrt{\hbar}\hat{\phi}$
and expand to O$\hat\phi^2$ 
The result is analogous to the Einstein case, but includes the $-\alpha \phi^2 R/12 $
term and becomes to order $\hbar$:
\bea 
\label{Jor0}
&&\backo
S_{Jordan}
= S_{Jc}+\hbar S_{Jq},
\nonumbo\backo
S_{Jc}=\int \sqrt{-g} \biggl(\frac{1}{2} \partial {\phi}_0 \partial{\phi}_0  
-V_J-\frac{\alpha}{12}\phi_0^{2}R+\frac{1}{2}M^{2}R
\biggr), 
\nonumbo\backo
S_{Jq}=\int \sqrt{-g} \biggl( \frac{1}{2} \partial \hat{\phi} \partial\hat{\phi}
-\half B_J \hat\phi^2 -\frac{\alpha}{12}(\hat\phi^2)R
\biggr),
\eea
where,
\bea &&
B_J
={3\lambda_1}\phi_0^{2}
+\frac{5\lambda_{2}}{2M^2}\phi_0^4 
\nonumbo
\qquad
+\frac{\lambda_3}{6M^2}
\biggl( 3\lambda_1\phi_0 ^{4}+(\partial\phi_0)^2
-\half\partial^2(\phi^2_0) \biggr),
\nonumbo
V_{J}=\frac{\lambda _{1}}{4}\phi_0^{4}
+\frac{\lambda_{2}}{12M^{2}}\phi_0 ^{6}
+\frac{\lambda_3}{12M^2}\phi_0 ^{2} (\partial \phi_0)^2.
\eea
Expanding to linear terms in $h_{\mu\nu}/M$  in weak field gravity 
the action eq.(\ref{Jor0}) becomes
\bea
\label{A4}&&
S_{Jq}\rightarrow \int \biggl( \half  \partial_{\mu }\hat\phi\partial^{\mu }\hat\phi
+\frac{h}{4M}\eta^{\mu\nu}\partial_{\mu }\hat\phi\partial_{\nu }\hat\phi
-\frac{h^{\mu \upsilon }}{2M}  \partial_{\mu }\hat\phi\partial _{\nu }\hat\phi
\nonumbo 
-\frac{\alpha}{12}R{\hat\phi^{2}}
- \half B_J\hat\phi^{2}-B_J\frac{h}{4M}\hat\phi^{2}
 \biggr).
\eea

Neglecting the contact terms we see, in addition to the non-curvature potential
obtained previously in eq.(\ref{20}),
the action eq.(\ref{A4}) now (naively) generates three diagrams linear in the curvature,
 D1, D2, and D3 of {Figs.(2,3,4)}. 
The additional D3 diagram (which is absent in the Einstein frame) is:
\bea &&
\label{D3}
\backo\backo
\Gamma_{D3}=
i \langle\;T \; 
 (-i\half  A\hat\phi^{2}R )
\;\;  (-i\half B_J\hat\phi^{2} ) \rangle
\nonumbo
=- AB_JR L =-\frac{\alpha}{6} B_JRL +{\cal{O}}\frac{R}{M^2}
\eea
{where $A=\alpha/6$}. 
Hence we have from eqs.(\ref{D1D2},\ref{D3}) for (D1+D2+D3):
\bea &&
\Gamma_{D3}+\Gamma_\alpha
= -\frac{1}{6}( \alpha-1 )  B_JR\;L 
\nonumbo
= 
-\frac{1}{6}( \alpha-1)
\biggl( {3\lambda_1}\phi_0^{2}
+{\cal{O}}\frac{1}{M^2}\biggr) RL
\eea
\begin{figure}[t]
\vspace{0 in}
	\hspace*{0.1in}\includegraphics[width=0.5\textwidth]{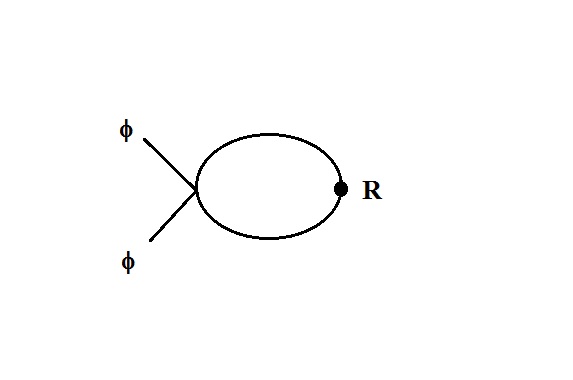}
	\vspace{-0.5 in}
	\caption{ Diagram D3 }
\end{figure}
Combining all effects at this point we have the
potential:
\bea &&
\label{pot19}
\backo
\Gamma_{Jordan}\equiv \frac{\lambda _{1}}{4}\phi_0^{4}
+\frac{\lambda_{2}}{12M^{2}}\phi_0 ^{6}
+\frac{\lambda_3}{12M^2}\phi_0 ^{2} (\partial \phi_0)^2
\nonumbo
\backo
- \biggl(
\frac{9\lambda^2_1}{2} \phi_0^{4} +\frac{15{\lambda_1\lambda_2 }+3\lambda^2_1\lambda_3}{2M^{2}}\phi_0^6
 +
\frac{3\lambda_3\lambda_1}{2M^2}{\phi_0^2}(\partial\phi_0)^2
\biggr) L
\nonumbo
\backo
-\frac{1}{6}( \alpha-1)
\biggl( {3\lambda_1}\phi_0^{2}\biggr) RL+\frac{\alpha}{12}\phi_0^{2}R
\eea
 This then contains the radiative correction and renormalization group running
 of $\alpha$
 \bea &&
 \alpha \rightarrow \alpha+6\lambda_1( 1-\alpha ) L
 \nonumbo
 D\alpha = -6\lambda_1( 1-\alpha )
 \eea
 To compare to the Einstein frame 
we now implement the contact terms directly from the action,
\bea && CT=
\int d^4x\bigg(-\frac{3F\partial ^{2}F}{4M^{2}}
+\frac{FT\left( \phi _{i}\right)}{2M^{2}}\bigg) 
\eea
where
\bea 
\label{R11}&&
 T=-(\partial\phi_0)^2+(\lambda_1 -18\lambda^2_1L) \phi_0^{4} 
 \nonumbo
 F=-\frac{\alpha}{6}\phi_0^2+( \alpha-1){\lambda_1}\phi_0^{2} L
 \eea
which leads to the contact term correction in the potential 
\bea &&\backo
-CT=-\frac{1}{12M^2}\biggl((\alpha+\alpha^2)\phi_0^2(\partial\phi_0)^2 -{\alpha}\lambda_1\phi_0^{6}
\nonumbo
\qquad
+\,(6\alpha+6 -12{\alpha^2}) \,{\lambda_1}\phi_0^{2}(\partial\phi_0)^2L
\nonumbo
\qquad
 + \, 6 ( 4\alpha-1){\lambda^2_1}\phi_0^{6} L \biggl)
\eea
and combining all effects the potential becomes:
\bea &&
\label{pot20}
\backo\Gamma = \frac{\lambda _{1}}{4}\phi_0^{4}
+\frac{\lambda_{2}}{12M^{2}}\phi_0 ^{6}
+\frac{\lambda_3}{12M^2}\phi_0 ^{2} (\partial \phi_0)^2
\nonumbo
\backo
- \biggl(
\frac{9\lambda^2_1}{2} \phi_0^{4} +\frac{15{\lambda_1\lambda_2 }+3\lambda^2_1\lambda_3}{2M^{2}}\phi_0^6
 +
\frac{3\lambda_3\lambda_1}{2M^2}{\phi_0^2}(\partial\phi_0)^2
\biggr) L
\nonumbo
\backo
-\frac{1}{12M^2}\biggl((\alpha+\alpha^2)\phi_0^2(\partial\phi_0)^2 -{\alpha}\lambda_1\phi_0^{6}
\nonumbo
\backo
+(6\alpha+6 -12{\alpha^2}) {\lambda_1}\phi_0^{2}(\partial\phi_0)^2L
 + ( 24\alpha-6){\lambda^2_1}\phi_0^{6} L\biggr)
\nonumbo
 \eea

Converting to the  Einstein frame variables:
\bea \label{ED1}
&&
\lambda_3' =\lambda_3 -\alpha-\alpha^{2}
\nonumbo
\lambda_1' = \lambda_1 
\nonumbo
\lambda_2'=\lambda_2+\alpha\lambda_1
\eea
we obtain after a somewhat tedious calculation,
  \bea
  \Gamma=\Gamma_E
+\biggl(
\frac{(8-3\alpha)\alpha   }{2M^{2}}\lambda'^2_1\phi_0^6
 -
{\frac{(4 +\alpha)\alpha }{2M^2}}\lambda'_1\phi_0^2(\partial\phi_0)^2
\biggr) L\nonumber
\eea
where $\Gamma_E(\phi_0)$ is given in eq.(61).
  
Comparing to eq.(\ref{all}) we therefore see an inconsistency between the { actions}
 $S_{Einstein}$ and  $S_{Jordan}$ at O($\hbar$) (opposite sign for potentials).
Hence we obtain the ``frame anomaly,'' 
\bea &&
\label{diff}
\backo
  S_{Jordan}- S_{Einstein}=
\nonumbo
\backo
-
{\int}\frac{(8
-3\alpha)\alpha   }{2M^{2}}\lambda'^2_1\phi_0^6L
 +
 {\int}
 {\frac{(4+\alpha)\alpha }{2M^2}}\lambda'_1\phi_0^2(\partial\phi_0)^2L
\eea
 We emphasize that the {\it rhs} of eq.(\ref{diff}) represents the mistake of
 not including the contact term in the initial action of eq.(\ref{Jor0}).
 For non-vanishing $\alpha$, the quantum  actions obtained in the two
   approaches agree if the quartic  scalar interaction is absent
   ($\lambda_1=0$). 

\section{CONCLUSIONS}

The Weyl transformation acting on the Jordan frame, to remove non-minimal interactions,
leading to the minimal Einstein frame, is identical to implementing
the contact terms \cite{HillRoss}.
If one didn't know about the Weyl transformation
one might discover it in the induced contact
terms in the single graviton exchange potential involving non-minimal couplings. 
The Weyl transformation is  powerful as it is fully non-perturbative.
Technically it can provide a useful check on the normalization and implementation
of the graviton propagators in various gauges.  But the contact term stipulates that the
mapping to the Einstein frame is dynamical and inevitable,
and does not involve field redefinitions.  Hence
there is no Jacobian in the path integral associated with going to the Einstein frame action
and it uniquely describes the theory.

In a model with non-minimal coupling $-\alpha\phi^2R/12$ this implies that the 
parameter $\alpha$ does not exist physically, unless demanded by a symmetry such as
  Weyl invariance.  Computing $\beta$-functions in a Jordan frame
without implementing the contact term will yield  incorrect results. 
 Implementing the contact term
yields the Einstein frame and results computed there will have no contact term ambiguities.

Nonetheless, Einstein frame will have a loop
induced infinitesimal $\alpha$ which can then be absorbed back into the potential terms
of the Einstein frame
by the contact terms (equivalently, a mini-Weyl transformation, 
or use of the modified $R$ equation of motion). 
The use of the modified $R$ equation of motion
on the non-minimal term is analogous to the use of the gluon field
equation for the electroweak penguin. It is likely that the Deans and Dixon \cite{Deans}
constraints on null operators apply to gravity as well.  

We emphasize that  {\it  our analysis applies strictly to
a theory with a Planck mass}.
A Weyl invariant theory, where  $M=0$, is nonperturbative
and our analysis is then inapplicable, and the Jordan frame is then
 physically relevant. Indeed, there is
no conventional gravity in this limit since the usual $M^2R$  (Fierz-Pauli) graviton kinetic 
term does not then exist.  Hence in this limit  one 
would have to appeal to a UV completion, e.g., string theory, $R^2$ gravity,
Weyl conformal geometry, etc.

 Contact term effects will disappear if we can go ``on-shell.''  
 This is demonstrated by the approach
 of Ruf and Steinwachs \cite{Ruf}, which employs an on-shell 
 calculational procedure.  However, the calculation
 of $\beta$-functions and effective actions is intrinsically an off-shell problem,
 since an  action is generally a functional of  fields that are unconstrained
 by equations of motion. The contact terms must be implemented for consistency.

In the case of an $R^2$ UV completion theory we  view the formation of
the Planck mass by, e.g., inertial symmetry breaking, i.e., as a dynamical
phase transition, similar to a disorder-order phase transition in
a material medium \cite{FHR}. 
It is interesting that in a $R_{\mu\nu}R^{\mu\nu}$ UV completion of gravity
such as in ref.\cite{Stelle}, 
the propagator becomes $1/q^4$,  nonetheless 
the contact term effective interactions exists above the Planck scale
for fields that couple non-minimally.  These then become
$q^2\times (1/q^4)\sim 1/q^2$.  Given the sign of $F(\phi_i)$ in eq.(\ref{action00}) there may exist an
inverse square law, pseudo-gravitational force that can be repulsive.
  This is one of many issues to develop further in
this context.

\bigskip
\noindent
 {\bf Acknowledgements}
\vspace{0.1in}

We dedicate this paper to our friend and colleague, Graham G. Ross.
This work grew out of previous collaborations with him, and we have benefitted
greatly from past discussions with him on this and other topics. 
Part of this work was done at 
Fermilab, operated by Fermi Research Alliance, 
LLC under Contract No. DE-AC02-07CH11359 with the United States 
Department of Energy. 
The work of D.G. was supported by a research grant PCE-2020-2255
of the Ministry of Education and Research.

%\newpage
\appendix
\section{Projection-Regulated Feynman Loops}

The loop induced effective potential for $\phi_0$ provides a useful way
to extract all of the $\beta$-functions of the various coupling constants.
The potential $\Gamma(\phi_0)$ is the log of the path integral:  $\Gamma=i\ln P$.
In the case of  a real scalar field with mass term
we consider the free action,
\bea &&
\half \int d^4x  \biggl( \partial\phi\partial\phi -m^2\phi^2\biggr)
\eea
we have for the path integral:
\bea
P=\underset{k}{\prod} \biggl(  k^{2}-m^{2} \biggr)^{-1/2}  
=\det \biggl( k^{2}-m^{2} \biggr)^{-1/2}
\eea
where $k=(k_0, \vec{k})$ is the $4$-momentum
hence, we have:
\bea
\label{A2}
\Gamma=i\ln P= -\frac{i}{2}  \int\frac{d^{4}k}{(  2\pi)  ^{4}}%
\ln \biggl(  k^{2}-m^{2}+i\epsilon \biggr)  
\eea
This can be evaluated with a Wick rotation to a Euclidean momentum, $k\rightarrow k_E=(ik_0,\vec{k})$, and a Euclidean
momentum space cut off $\Lambda$: 
\bea
\label{A3}
\Gamma &= & \frac{1}{2}\int_0^{\Lambda}\frac{d^{4}k_E}{ (  2\pi )^{4}}
\ln \biggl(  \frac{k_E^{2}+m^{2}}{\Lambda^{2}} \biggr) 
\nonumber \\
&&
\backo =
\frac{1}{64\pi^{2}} \biggl( 
\Lambda^{4} \ln\frac{\Lambda^{2}+m^{2}}{\Lambda^{2}}
-m^{4}\ln\frac{\Lambda^{2}+m^{2}}{m^{2}}
\nonumbo
\backo\backo\qquad
-\frac{1}{2} \Lambda^{4}
+\Lambda^{2}m^{2} \biggr)  + (\makebox{irrelevant constants})
\eea
where we inserted $\Lambda^{-2}$ in 
the arguments of the logs to preserve zero scale dimension.
The
cutoff can be viewed as a spurious parameter, introduced to make the integral finite
and arguments of logs dimensionless, but it is not part of the defining
action. 
The only physically meaningful dependence upon $\Lambda$ is contained 
in the
logarithm, where it reflects scale symmetry breaking by the 
quantum trace anomaly. Powers of $\Lambda$, e.g., $\Lambda^{4},\Lambda^{2}m^{2}$, spuriously break classical scale
symmetry and are not part of the classical action  \cite{Bardeen}.

It is therefore conceptually useful to have a definition of the
loops in which the spurious powers of $\Lambda$ do not arise.
This can be done by defining the loops
applying projection
operators on the integrals.  The projection operator
\bea
P_{n}= \biggl(  1-\frac{\Lambda}{n}\frac{\partial}{\partial\Lambda} \biggr)  
\eea
removes any terms proportional to $\Lambda^{n}.$ 
Since the defining classical Lagrangian
has mass dimension 4 and involves no terms with  $\Lambda^{2}m^{2}$ or
$\Lambda^{4},$ we define the regularized loop integrals as:
\bea
\label{19}
\Gamma &\rightarrow &\frac{1}{2}P_{2}P_{4}\int_0^{\Lambda}\frac{d^{4}k_E}{ (  2\pi )^{4}}
\ln \biggl(  \frac{k_E^{2}+m^{2}}{\Lambda^{2}} \biggr) 
\nonumber \\
&=&
-\frac{1}{64\pi^{2}}  m^{4} \biggl(  \ln\frac{\Lambda^{2}}{m^{2}%
}\biggr)   +O \biggl(  \frac{m^{6}}{\Lambda^{2}} \biggr)    
\eea
where we then take the limit $\Lambda\gg m$ to suppress $O ( m^{6}/
{\Lambda^{2}})  $ terms and we are interested 
only in the log term (not additive constants)
 This means that the additive, non-log terms, e.g.
$c'm^2$, are undetermined, and the only physically meaningful  result is the 
$\ln(\Lambda^2/m^2)$ term. $\Lambda $ can be swapped for a running renormalization scale $\mu$.
Interestingly, if we define the integral as $\int dX\rightarrow P_1P_2P_3...P_\infty\int dX$,
the action on the logs will lead to the Euler constant that arises in dimensional regularization,
hinting at a mapping to the dimensionally regularized result.

\end{document}